\documentclass[prb,twocolumn,preprintnumbers,amsmath,amssymb,english,showpacs,superscriptaddress,floatfix]{revtex4}
\usepackage{mathptmx}
\usepackage{helvet}
\usepackage[T1]{fontenc}
\usepackage[latin1]{inputenc}
\usepackage{amsthm}
\usepackage{graphicx}
\usepackage{babel}
\usepackage{natbib}
\begin{document}

\title{Photo-absorption spectra of small hydrogenated silicon clusters using
the time-dependent density functional theory}
\author{Juzar Thingna}
\affiliation{Physics Department, Center for Computation Science and Engineering,
National University of Singapore, Singapore}
 \affiliation{Physics Department, Indian Institute of Technology Kanpur, Kanpur
UP 208016, India }
\author{R. Prasad}
 \email[Electronic Address:]{rprasad@iitk.ac.in}
\author{S. Auluck}
  \affiliation{Physics Department, Indian Institute of Technology Kanpur, Kanpur
UP 208016, India }

\date{\today}
\begin{abstract}
We present a systematic study of the photo-absorption spectra of various
Si$_{n}$H$_{m}$ clusters (n=1-10, m=1-14) using the
time-dependent density functional theory (TDDFT). The method uses
a real-time, real-space implementation of TDDFT involving full propagation
of the time dependent Kohn-Sham equations. Our results for SiH$_{4}$
and Si$_{2}$H$_{6}$ show good agreement with the earlier calculations
and experimental data. We find that for small clusters (n<7) the
photo-absorption spectrum is atomic-like while for the larger clusters
it shows bulk-like behaviour. We study the photo-absorption spectra
of silicon clusters as a function of hydrogenation. For single hydrogenation,
we find that in general, the absorption optical gap decreases and
as the number of silicon atoms increase the effect of a single hydrogen
atom on the optical gap diminishes. For further hydrogenation the
optical gap increases and for the fully hydrogenated clusters the
optical gap is larger compared to corresponding pure silicon clusters.

PACS number(s): 78.67. -n, 73.22.-f, 71.15 Pd
\end{abstract}
\maketitle

\section{Introduction}
Recently, there has been a renewed interest in understanding the optical
properties of clusters because confinement of electrons changes the
physical properties. Thus by varying size of the clusters the optical
properties can be tuned according to the desired application\cite{Zdetsis}.
In particular, optical properties of silicon and hydrogenated silicon
clusters have been of great interest due to the observation of photo luminescence
(PL) in porous silicon \cite{Canham,Dele}. The structure and properties
of silicon clusters can be tuned by varying the cluster size as well
as doping. An important dopant for silicon clusters is hydrogen and
it plays an important role in structural stability. Experimental studies
also confirm this fact\cite{Mura}, but in spite of several investigations
many issues about hydrogenated silicon clusters have not been understood.
It is not clear how the structure and optical properties of the cluster
evolves with size and as a function of hydrogenation. It is therefore
interesting to calculate the photo-absorption (PA) spectra and compare
it with experiment.

There has been a considerable amount of work experimental as well
as theoretical on silicon and hydrogenated silicon clusters. Suto
and Lee measured the photo-absorption and fluorescence of silane in
the energy range 8-12 eV using synchrotron radiation \cite{Suto}.The
optical absorption of silane and disilane has been measured by Itoh
et. al. in the energy range 6-12 eV\cite{Itoh}. Cheshnovsky et. al.
have measured the photo-electron spectra of charged silicon and germanium
clusters\cite{Ches}. Rinnen and Mandich have measured the photo-dissociation
spectra of neutral silicon clusters Si$_{n}$(n=18-41)\cite{Rinn}.
Murakami and Kanayama investigated the stability of some silicon and
hydrogenated silicon clusters using a quadruple ion trap\cite{Mura}.
More recently Antonietti et. al. deduced the photo-absorption spectra
of charged silicon clusters from photo-dissociation of charged xenon-silicon
clusters\cite{Anto}. On the theoretical side Chantranupong et. al.
performed a configuration interaction (CI) calculation for a large
number of low lying states in silane\cite{Chan}. Rubio et. al. calculated
the photo-absorption spectra of silicon and alkali metal clusters
using time dependent local density approximation (TDLDA)\cite{Rubi}.
Rohlfing and Louie calculated the optical absorption spectra of hydrogen
terminated silicon clusters by solving the Bethe-Salpeter equation\cite{Rohl}.
Vasiliev et. al. \cite{Vasi} and Marques et. al. \cite{Mar3}calculated
the optical absorption spectra of Si$_{n}$H$_{m}$ clusters
using linear response theory within TDLDA. Rao et. al. measured the
photo luminescence of a dispersion of 1 nm silicon particles obtained
from crystalline silicon that is dispersed into nanoparticles through
electrochemical etching with HF and H$_{2}$O$_{2}$\cite{Rao}.
They also calculated the photo-absorption spectra of Si$_{29}$H$_{24}$
using time dependent density functional theory (TDDFT). Lehtonen and 
Sundholm calculated the absorption spectra of three hydrogen terminated
silicon clusters using TDDFT\cite{Leht}.

The earlier studies on the optical properties of silicon and hydrogenated
silicon clusters focused on the size dependence of the PL and photo-absorption
\cite{Taka,Hill,Dele,Wang,Ogut,Dell}. Some of these studies ignored
the influence of oscillator strength's (electric dipole matrix elements)
and hence were not in good agreement with the experimental data. Calculations
based on DFT using LDA and the generalised gradient approximation
(GGA) which included the dipole matrix elements suffered from the
drawback that is inherent in LDA/GGA i.e. the energy gaps were underestimated
because these calculations ignored the effect of excited states. To
overcome this drawback, the configuration interaction method or the
methods based on solving the Bethe-Salpeter equation along with the
GW approximation have been suggested\cite{Rohl}. These methods require
a lot of computer time and hence have been restricted to small clusters.
A computational technique based on linear response theory within TDLDA
had been proposed by Vasiliev et. al.\cite{Vasi}. This is a natural
extension of the LDA ground state density functional formulation designed
to include excited states. This method is faster than the BS and GW
methods and can therefore be used for larger clusters. Vasiliev et.
al. have shown the viability of their method by performing calculations
on silicon and hydrogenated and oxygenated silicon clusters. Another
implementation of TDDFT has been formulated by Castro et. al. which
allows the calculation of the excited state energies and optical absorption
spectra. In the present work, we have used this method\cite{Cast}. 

Most of earlier calculations have considered only hydrogen terminated
silicon clusters. There seems to be a lack of a systematic study on
small hydrogenated silicon clusters. In this paper we report calculations
for the silicon and hydrogenated silicon clusters in addition to some
hydrogen terminated silicon clusters. Thus our calculations will show
the effect of hydrogenation on the optical properties of silicon clusters.
Our emphasis is on the smaller clusters so as to bring out the evolution
of the optical properties as we increase the number of silicon and
hydrogen atoms in the cluster.

The plan of the paper is as follows. In section II, we briefly discuss
the method and give computational details. In section III we present
our results and discussion. In section IV we present our conclusions.

\section{Method and Computational Details}
We have used TDDFT for our calculation of the photo-absorption spectrum\cite{Mar1,Mar2}.
For the sake of completeness we shall summarise the essentials of
the method. In TDDFT, the basic variable is the one electron density
$n(\mathbf{r},t)$, which is obtained with the help of a fictitious
system of non-interacting electrons, the Kohn-Sham system. The time-dependent
Kohn-Sham equations are
\begin{equation}
i\frac{\partial}{\partial t}\psi_{i}(\mathbf{r},t)=\left[-\nabla^{2}/2+v_{KS}(\mathbf{r},t)\right]\psi_{i}(\mathbf{r},t)
\end{equation}
where $\psi_{i}(\mathbf{r},t)$ are Kohn-Sham one electron orbitals.
In terms these orbitals $n(\mathbf{r},t)$ can be written as \begin{equation}
n(\mathbf{r},t)=\sum_{i}^{occ}|\psi_{i}(\mathbf{r},t)|^{2}\end{equation}
The Kohn-Sham potential can be written as 
\begin{equation}
v_{KS}(\mathbf{r},t)=v_{ext}(\mathbf{r},t)+v_{Hartree}(\mathbf{r},t)+v_{xc}(\mathbf{r},t)
\end{equation}
where the first term is the external potential, the second Hartree
potential and the last the exchange and correlation potential.
For obtaining this potential we use adiabatic local density approximation
(ALDA).

In our work we shall use the TDDFT scheme which calculates propagation
of the time-dependent Kohn-Sham equations in real time. In this scheme
the electrons are given some small momentum($\kappa$) to excite all
the frequencies\cite{Yaba}. This is achieved by transforming the
ground state wave function according to
\begin{equation}
\psi_{i}(\mathnormal{\mathbf{r}},\delta t)=e^{i\kappa z}\psi_{i}(\mathbf{r},0)
\end{equation}
and then propagating these wave functions for some finite time. The
spectrum is then obtained from dipole strength function $S(\omega)$
\begin{equation}
S(\omega)=\frac{2\omega}{\pi}Im\alpha(\omega)
\end{equation}
where the $\alpha(\omega)$is the dynamical polarizability and is
given by 
\begin{equation}
\alpha(\omega)=\frac{1}{\kappa}\int dte^{i\omega t}[d(t)-d(0)]
\end{equation}
where $d(t)$ is dipole moment of the system.

We can also define a quantity know as the oscillator strength to express
the strength of the transition. 
\begin{equation}
f_{I}(x)=\frac{2}{3}\omega_{I}\sum_{n\in x,y,z}|<\phi_{0}|n|\phi_{I}>|^{2}
\end{equation}
where $\phi_{0}$ and $\phi_{I}$ are the ground and excited state
respectively. The oscillator strength is related to the dipole strength
function defined earlier using the following relationship
\begin{equation}
S(\omega)=\sum_{I}f_{I}\times\delta(\omega-\omega_{I})
\end{equation}
The sum over the oscillator strength gives the number off active electrons
in the system,
\begin{equation}
\sum f_{I}=N
\end{equation}
where \textit{N} is the number of active electrons in the system.

The initial structures of the clusters used for our calculations have
been obtained by Balamurugan and Prasad\cite{Bal1,Bal2} earlier from
their analysis based on the Car-Parrinello molecular dynamics (CPMD)
\cite{CPMD}. The resulting structures have been further optimised
using the electronic structure method implemented in VASP \emph{(Vienna
Ab-initio Simulation Package})\cite{Kres}. We do not find any significant
changes in the structures compared to the CPMD results. In fact the
starting CPMD configuration gives forces that are around 0.5 eV/\AA\ which eventually reduce to around 0.05 eV/\AA\ in VASP. 
VASP employs density functional theory (DFT) and we have
employed the local density approximation (LDA) for the exchange correlation
using norm conserving pseudo-potentials. The optimisation is done
only by relaxing the ions via the conjugate gradient (CG) method and
using a k-point Monkhorst-Pack mesh of 4x4x4. All calculations have
been performed in a cubic supercell of length 20 \AA.

In this work the photo-absorption spectra of the optimised structures
was studied using OCTOPUS code \cite{Cast,Marq} where the above approach
is implemented. All calculations are expanded in a regular mesh in
real space, and the simulations are performed in real time. The local
density approximation was employed to keep consistency with the geometry
optimisation process. OCTOPUS uses a uniform grid in real space, which
is located inside the sum of n spheres, one around each atom of the
n-atom cluster. For all clusters a minimisation of energy with respect
to the radius and the grid spacing was carried out. We required the
radius of each sphere to be 6-8 \AA\ and the
grid spacing 0.28-0.4 \AA\ for optimal energy
minimisation.

For the TDDFT the propagation in real time has been performed with
30,000 time steps with a total simulation time of around 124 fs. This
gives a good resolution of the spectra. Throughout the calculations
the ions were kept static and in order to approximate the evolution
operator the approximated enforced time reversal symmetry (aetrs)
method was employed. Numerically the exponential of the Hamiltonian,
which is used to approximate the evolution operator was evaluated
using a simple Taylor expansion of the exponential.

\begin{table*}
\centering{}\caption{Excitation energies in eV for SiH$_{4}$and Si$_{2}$H$_{6}$ clusters.}
\begin{tabular}{|c|c|c|c|c|c|c|}
\hline 
 & Transition  & Present Work & Marques et. al. & Rohlfing \& Louie  & Vasiliev et. al.  & Experiment \tabularnewline
\hline
\textbf{SiH$_{4}$}  & 4s  & 8.2  & 8.2 & 9.0  & 8.2  & 8.8$^{\text{a}}$,9.0$^{\text{b}}$ \tabularnewline
 & 4p  & 9.2  & 9.4 & 10.2  & 9.2  & 9.7$^{\text{a,b}}$ \tabularnewline
 & 4d  & 9.8  & 10 & 11.2  & 9.7  & 10.7$^{\text{a,b}}$ \tabularnewline
\textbf{Si$_{2}$H$_{6}$}  & 4s  & 7.3  & 7.3 & 7.6  & 7.3  & 7.6$^{\text{a}}$ \tabularnewline
 & 4p  & 8.3  & - & 9.0  & 7.8  & 8.4 $^{\text{a}}$\tabularnewline
 &  & 8.6  & 8.7 &  &  & 9.5$^{\text{a}}$ \tabularnewline
 &  & 10.6  & 10.6 &  &  & 9.9 $^{\text{a}}$\tabularnewline
\hline
\end{tabular}
\end{table*}

\section{Results and Discussions}
In order to verify our calculations with the experimental data, we
chose to compare the two most stable clusters SiH$_{4}$ and Si$_{2}$H$_{6}$
with the available experimental data as our benchmark. In optical
properties we are interested in transitions form the occupied levels
to the unoccupied levels. The structures in the photo-absorption spectra
are identified with these transitions. It is therefore interesting
to calculate these energy differences and compare them with the energy
differences deduced from the experimental spectra. Table I shows such
a comparison. For our calculations, we have checked that the f-sum
rule is satisfied\cite{Woot}. In Table 1 we have also included the
Bethe-Salpeter results of Rohlfing and Louie \cite{Rohl}, the TDLDA
results of Vasilev et. al.\cite{Vasi} and the TDLDA results from
Octopus of Marques et. al.\cite{Mar3} along with the transitions
identification. The uppermost occupied states result from a hybridisation
of the silicon and hydrogen states while the lowest unoccupied states
are primarily silicon states. Thus the energy gap is dependent on
the the bonding and anti-bonding silicon states. The transition between
these states have been identified as 4s, 4p, 4d states (these refer
to the angular momentum character of the final states). We find good
agreement with the experimental data of Itoh et. al. \cite{Itoh}and
Suto and Lee\cite{Suto} and other theoretical calculations. In particular,
we get very good agreement with the results of Marques et. al.\cite{Mar3}.
As compared to the work of Marques et. al. we obtain an additional
peak at 8.3 eV for Si$_{2}$H$_{6}$ which is observed in other works
\cite{Vasi} and the experiments\cite{Itoh,Suto}. Our simulations
pick up this additional peak because our broadening factor, which
is the inverse of the time steps in the TDDFT run is quite narrow
as compared to Marques et. al. Fig. 1 shows the structures and photo-absorption
spectra of the current work, Vasiliev et.al. \cite{Vasi} and Marques
et. al. \cite{Mar3} for SiH$_{4}$ and Si$_{2}$H$_{6}$ clusters.
The calculated photo-absorption cross-section (using TDDFT) seems
to be in good agreement with the TDLDA results of Vasiliev et. al.
\cite{Vasi} and the TDLDA results of Marques et. al. \cite{Mar3}
using Octopus.

\begin{figure}
\caption{Structure and photo-absorption spectra of present work (Black solid line),
Vasiliev et.al. \cite{Vasi} (Blue dashed line) and Marques et.
al. \cite{Mar3} (Red solid line) for SiH$_{4}$ and Si$_{2}$H$_{6}$
clusters.}
\includegraphics[scale=0.3]{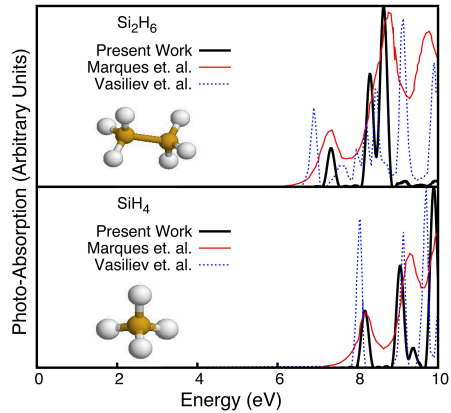}
\end{figure}

In order to see the effect of increasing the size of silicon clusters
on the optical spectra, we present in Fig. 2 I the optimised structures
of silicon clusters Si$_{n}$ [n=1-10] along with the
photo-absorption spectra. To see the effect of adding a single H atom
to these clusters, we have also calculated the photo-absorption spectra
of Si$_{n}$H clusters which are also shown in Fig. 2 II along
with the silicon clusters results. Each figure shows the structure
of silicon and hydrogenated silicon clusters and the calculated photo-absorption
cross-section. Consider first the silicon clusters. For small clusters
(up to n=7) we find that the photo-absorption spectra is a combination
of many peaks and looks like that of isolated atoms. However for n>7, 
the optical spectra looks bulk-like\cite{Adol}. This can be
understood using a simple tight-binding picture. In a larger silicon
cluster, the overlap between electronic wave-functions lifts the degeneracy
of the energy levels resulting in bunching of energy levels in a narrow
energy range. This results in broadening of PA spectrum for larger
clusters. In Fig 2 we see that the main structure in the PA spectra
is located at around 9 eV and a minor structure starts to build up
at around 15 eV. The same trends are found in the singly hydrogenated
clusters. For smaller clusters (n<7) the PA spectrum changes significantly
as we increase the number of hydrogen atoms, while for larger clusters
this change is small.

\begin{figure}
\caption{I: Structure and photo-absorption spectra of Si$_{n}$ clusters
(n=1-10 from top to bottom) ; II: Structure and photo-absorption spectra
of Si$_{n}$H clusters (n=1-10 from top to bottom).} \footnote{Solid line corresponds to the PA spectra of the ground state structure Fig. 4 (I), whereas dashed line corresponds to the PA spectra of higher energy state structure Fig. 4 (II).}
\includegraphics[scale=0.3]{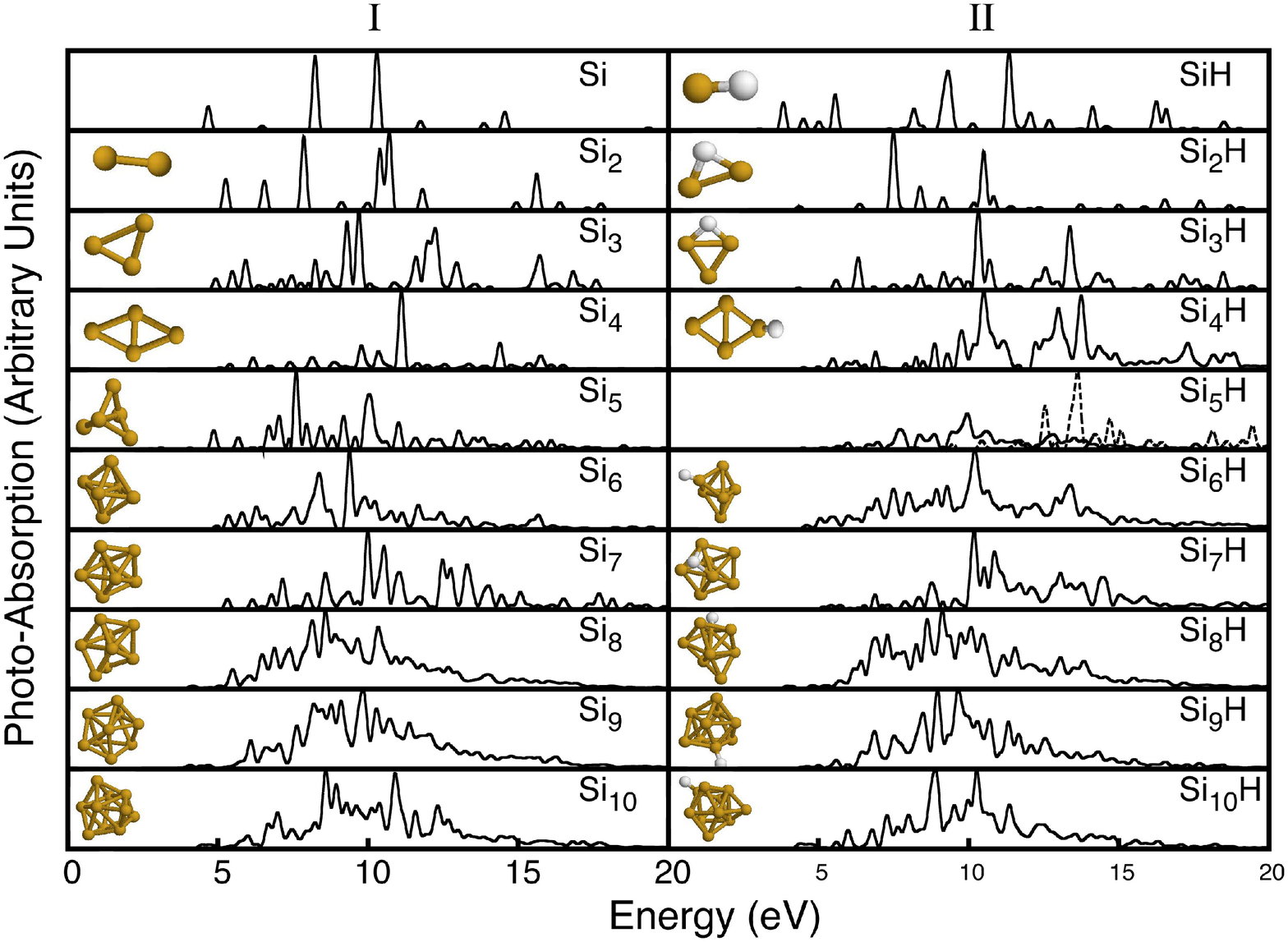}
\end{figure}

To see the effect of further hydrogenation, we present the PA spectra
of hydrogenated silicon clusters where the number of hydrogen atoms
is increased in Fig. 3. We have performed calculations for Si$_{3}$H$_{3}$,
Si$_{4}$H$_{4}$ and Si$_{4}$H$_{8}$,
Si$_{5}$H$_{6}$ and Si$_{5}$H$_{12}$
and Si$_{6}$H$_{7}$ and Si$_{6}$H$_{14}$
clusters. The different positions of the hydrogen atoms in Si$_{3}$H
and Si$_{3}$H$_{3}$ makes the two structures very
different. Hence there is a large difference between the PA spectra.
In going from Si$_{3}$H to Si$_{3}$H$_{3}$
we find that the number of structures in the PA spectra increases
and some of the new structures have larger peaks. In fact the PA spectrum
goes from atom-like to bulk-like. We see similar trends in Si$_{4}$H$_{4}$
and Si$_{4}$H$_{8}$ except that the silicon backbone
is a rhombus. We find that the new structures in Si$_{4}$H$_{8}$
PA spectrum have significantly larger peaks compared to Si$_{4}$H$_{4}$.
In these hydrogenated clusters (Si$_{3}$H$_{3}$, Si$_{4}$H$_{4}$, Si$_{4}$H$_{8}$, Si$_{5}$H$_{6}$
, Si$_{5}$H$_{12}$, Si$_{6}$H$_{7}$
and Si$_{6}$H$_{14}$) which have different silicon
backbones, we observe that the bulk-like behaviour seems to start at
n=4 which is much earlier than in singly hydrogenated and silicon
clusters. This effect may be attributed to the large number of hydrogen
atoms.

We obtained two very closely related structures of Si$_{5}$H having
a difference of about 0.043 eV in their total energy. Both the structures
have been shown in Fig. 4 and the corresponding PA spectra in Figure
2 II. The silicon backbone in both these clusters remains the same
and the only difference is found in the position of the hydrogen atom.
In the ground state structure (solid line in the PA spectra) hydrogen
is bonded to two silicon atoms while in the other (dashed line in
the PA spectra) it is bonded to only one silicon atom. The PA spectra
of the two structures is quite different. This shows that the PA spectrum
is sensitive to small changes in the structure. Thus the PA spectrum
can be used to identify the structure of the clusters which differ
by a very small energy. 

\begin{figure}
\caption{Structure and photo-absorption spectra of Si$_{n}$H$_{m}$
clusters (from top to bottom: Si$_{3}$H$_{3}$, Si$_{4}$H$_{4}$, Si$_{4}$H$_{8}$, Si$_{5}$H$_{6}$, Si$_{5}$H$_{12}$, Si$_{6}$H$_{7}$, Si$_{6}$H$_{14}$)}
\includegraphics[scale=0.3]{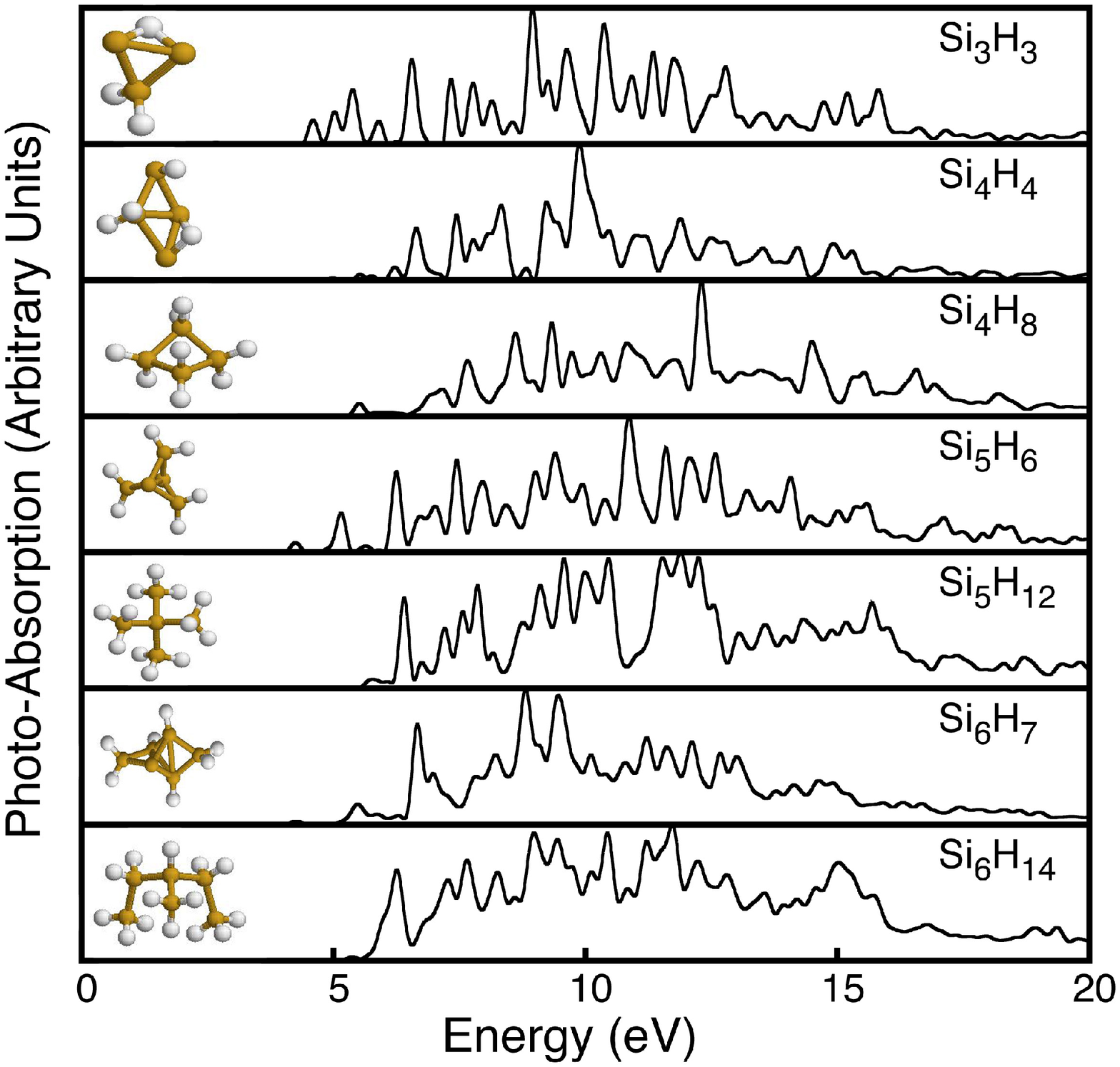}
\end{figure}

In Table II we present the optical absorption gaps of the various clusters.
We define the optical gaps through the integral oscillator strength
rather than as the energy of the first dipole allowed transition in
the absorption spectra. The integral oscillator strength gives the
total number of active electrons in the system. In this approach the
value of the optical absorption gap is determined at a very small
but non-zero fraction of the complete oscillator strength\cite{Vasi-1}.
We set this threshold to $10^{-4}$ of the total oscillator
strength. This value is chosen because it stands above the value of 
``numerical noise''and at the same time it is sufficiently small
so as to not suppress the experimentally detectable dipole allowed
transitions. This definition of absorption gap doesn't affect the
values of optical gaps for small clusters, since the intensity of
first transitions is well above the selected threshold. 
\begin{figure}
\caption{Structures of ground state (I) and higher energy state (II) Si$_{5}$H
clusters.}
\includegraphics[scale=0.3]{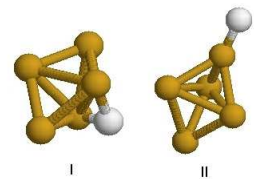}
\end{figure}

In Fig. 5 we plot the optical gap as a function of the number of silicon
atoms (n) for various Si$_{n}$ (solid Line) and Si$_{n}$H
(dashed Line) clusters (n=1-10). It shows a general trend that the
addition of a single hydrogen atom reduces the optical gap as compared
to the silicon clusters. As the number of silicon atoms in the cluster
increase we find that the effect of a single hydrogen atom reduces
the optical gap and the difference between the optical gap of Si$_{n}$
and Si$_{n}$H gradually decreases with the increase in silicon
atoms. In the bulk limit a single hydrogen atom should not distort
the optical gap and thus in that limit the optical gap of the silicon
will be the same as the optical gap of the singly hydrogenated silicon.

To see the effect of further hydrogenation, we list the optical gaps
of all the clusters in Table II. We see that addition of a single
hydrogen atom reduces the optical gap, in general, but with further
hydrogenation the optical gap increases. For the fully hydrogenated
clusters the optical gap is larger than the unhydrogenated clusters.
Thus in the bulk limit the hydrogenated system will show a larger
gap which is consistent with the experimental observations.\cite{Cody}

\begin{figure}
\caption{Optical absorption gap of various Si$_{n}$ (solid line filled circle)
and Si$_{n}$H (dashed line empty circle) clusters as a function of
the number of silicon atoms (n).}
\includegraphics[scale=0.3]{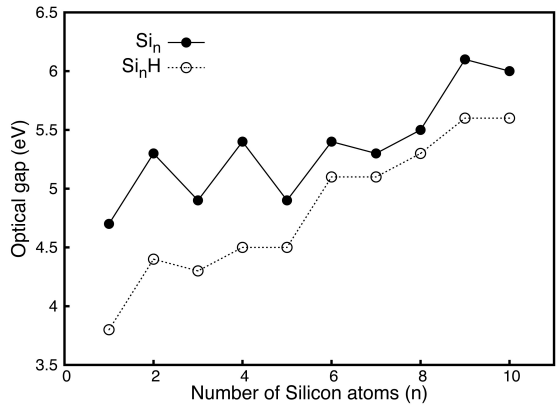}
\end{figure}

\section{Conclusions}

The photo-absorption spectra of silicon and hydrogenated silicon clusters
have been calculated using TDDFT. Our calculations show very good
agreement for the benchmarks of SiH$_{4}$ and Si$_{2}$H$_{6}$ with
the earlier theoretical calculations and experimental results. We
find that the changes in the photo-absorption spectra correlates well
with the changes in the structure of the cluster. In the singly hydrogenated
clusters we find bulk-like behaviour for larger clusters (n>7) while
for the smaller clusters we find the PA spectra is composed of numerous
peaks as in atoms. We find that the PA spectra of two structures of
Si$_{5}$H which are close in energy to be quite different.
For the other hydrogenated clusters the bulk-like behaviour appears
for smaller n which can be attributed to the larger number of hydrogen
atoms.The addition of a single hydrogen to the silicon cluster shows
a decrease in the absorption optical gap. The optical gap decreases
gradually with the number of silicon atoms in the cluster giving the
bulk limit in which an addition of a single hydrogen to the silicon
should not effect the optical gap of silicon. With further hydrogenation
the optical gap increases and for the fully clusters the optical gap
is larger than the corresponding gap for pure silicon clusters.

\section{Acknowledgement}

We are grateful to D Balamurugan for providing the co-ordinates of
some clusters. We thank Professor M. K. Harbola for helpful discussions.

\begin{table}
\caption{Optical gaps of various clusters.}

\centering{}\begin{tabular}{|c|c|}
\hline 
Cluster & Absorption Optical Gap (eV) \tabularnewline
\hline
Si & 4.7\tabularnewline
SiH & 3.8\tabularnewline
SiH$_{4}$ & 8.2\tabularnewline
\hline
Si$_{2}$ & 5.3\tabularnewline
Si$_{2}$H & 4.4\tabularnewline
Si$_{2}$H$_{6}$ & 7.3\tabularnewline
\hline
Si$_{3}$ & 4.9\tabularnewline
Si$_{3}$H & 4.3\tabularnewline
Si$_{3}$H$_{3}$ & 4.6\tabularnewline
\hline
Si$_{4}$ & 5.4\tabularnewline
Si$_{4}$H & 4.5\tabularnewline
Si$_{4}$H$_{4}$ & 5.5\tabularnewline
Si$_{4}$H$_{8}$ & 5.6\tabularnewline
\hline
Si$_{5}$ & 4.9\tabularnewline
Si$_{5}$H (gs) & 4.5\tabularnewline
Si$_{5}$H (he) & 5.4\tabularnewline
Si$_{5}$H$_{6}$ & 5.2\tabularnewline
Si$_{5}$H$_{12}$ & 6.4\tabularnewline
\hline
Si$_{6}$ & 5.4\tabularnewline
Si$_{6}$H & 5.1\tabularnewline
Si$_{6}$H$_{7}$ & 4.3\tabularnewline
Si$_{6}$H$_{14}$ & 6.3\tabularnewline
\hline
Si$_{7}$ & 5.3\tabularnewline
Si$_{7}$H & 5.1\tabularnewline
\hline
Si$_{8}$ & 5.5\tabularnewline
Si$_{8}$H & 5.3\tabularnewline
\hline
Si$_{9}$ & 6.1\tabularnewline
Si$_{9}$H & 5.6\tabularnewline
\hline
Si$_{10}$ & 6.0\tabularnewline
Si$_{10}$H & 5.6\tabularnewline
\hline
\end{tabular}
\end{table}

\clearpage
\newpage{}

\end{document}